\documentstyle[epsf,seceq]{ptptex}

\newcommand{\simgt}{\lower.5ex\hbox{$\; \buildrel > \over \sim \;$}}
\newcommand{\simlt}{\lower.5ex\hbox{$\; \buildrel < \over \sim \;$}}

\title{%
{\baselineskip0pt
\leftline{\large\baselineskip16pt\sl\vbox to0pt{\hbox{ } 
               \hbox{ }\vss}}
}\\
\vspace{1cm}
Feasibility of Probing Dark Energy with Strong Gravitational 
Lensing Systems}
\subtitle{Fisher-Matrix Approach}

\author{
Kazuhiro {\sc Yamamoto}${}^{1}$, 
Yasufumi {\sc Kadoya}${}^{1}$,
Tsukasa  {\sc Murata}${}^{1}$ and \\
Toshifumi {\sc Futamase}${}^{2}$ 
}

\inst{
${}^{1}$Department of Physical Science, Hiroshima University, 
\\
Higashi-Hiroshima 739-8526, Japan
\\
\vspace{1mm}
${}^{2}$Astronomical Institute, Graduate School of Science, 
Tohoku University,
\\ 
Sendai 980-8578, Japan }

\recdate{June 8, 2001}

\begin{document}

\abst{%
We assess the feasibility of probing dark energy with strong 
gravitational lensing systems. The capability of the method, 
which depends on the accuracy with which the lensing systems 
are modeled, is quantitatively investigated using the Fisher-matrix 
formalism. We show that this method might place useful constraints 
on the density parameter and the redshift evolution of the dark 
energy by combining it with a constraint from supernova measurements.
For this purpose, the lens potential needs to be precisely reconstructed.
We determine the required quality of data. 
We also briefly discuss the optimal strategy to constrain the 
cosmological parameters using gravitational lensing systems.
}

\maketitle
\section{\bf Introduction}
\def\F{{F}}
Recent observations of distant Type Ia supernovae have provided strong
evidence for the acceleration of the universe.\cite{Perlmuttereetal,Riessetal} 
Recent measurements of cosmic microwave background anisotropies 
favor a spatially flat universe with cold dark 
matter,\cite{Bernardis,Lange} 
while measurements of the large-scale structure in the distribution of 
galaxies favor a low-density universe.\cite{Peacock} 
These observations can be explained by the hypothesis that our universe 
is dominated by dark energy in addition to cold dark matter.
Motivated by these findings, many authors have recently investigated 
the origin of dark energy.\cite{rf:Caldwell98,Arkani,Parkerb}
An attractive feature of these theoretical models is that 
the coincidence problem,
the near coincidence of the density of matter and the dark 
energy component may be explained,\cite{Arkani,rf:Zlatev98,Parkera}
though the plausibility is under debate.\cite{Vilenkin}

An evolving scalar field has been investigated as a model of
a decaying cosmological constant (dark energy).\cite{PR,FN,FON,Fujii98}  
Dark energy can be characterized by the effective equation of 
state $w=\rho_D/P_D$, where $\rho_D$ and $P_D$ are the energy 
density and the pressure of the dark energy, respectively.
Different theoretical scenarios of dark energy may be distinguished 
by the measurement of the effective cosmic equation of state. For 
example, the quintessence model generally predicts that $w$ can 
be a function of redshift and satisfies $w\geq -1$.\cite{rf:Caldwell98}
If dark energy originated from the cosmological constant or a false 
vacuum energy, $w=-1$.\cite{Arkani} Parker and Raval have explained
dark energy in terms of a quantum vacuum effect produced by the curved 
spacetime of general relativistic cosmology. In their model, 
$w\leq-1$ is allowed.\cite{Parkerb} Models which predict $w\leq-1$
have also been proposed within the framework of the evolving 
scalar field.\cite{Phantom,COY}
Therefore, the measurement of $w$ may provide information regarding  
the origin of dark energy. 
For this reason various observational strategies have been proposed to 
probe characteristics of dark 
energy.\cite{rf:Efs,rf:Wang,rf:ND,CNS,NC99,NC01,HTa,rf:Saini,rf:chiba,YN}
One of the most promising approaches for determining $w$ 
employs type Ia supernovae, while other approaches using 
CMB anisotropies and number-counts should be useful too.\cite{HTb,SS,YS}
However, it has been pointed out that the method using 
supernovae is fundamentally limited because the luminosity distance 
depends on $w$ through a multiple integration, which smears out 
information concerning $w$ and its time variation.\cite{Maor}

With the situation as described above, it is very useful to have an independent
method to determine $w$. One possible such method is to use
gravitational lensing. In fact, it is known
that lensing statistics depend strongly on the 
value of the cosmological constant, as well 
as the characteristics of dark energy.\cite{Turner,FFKT,rf:FFK,rf:CH,rf:Zhu}
Another possible method to probe dark energy has been 
proposed.\cite{rf:FH,rf:FY,rf:YF,IGR} This method relies on precise 
measurements of `clean' gravitational lensing systems.
Here `clean' means suitable for modeling 
a gravitational lensing system, as is an Einstein ring or Einstein cross. 
Motivated by this recent proposition, we assess the feasibility 
of the method to probe dark energy that employs the Fisher-matrix formalism, 
which is useful to demonstrate how accurately one can estimate model 
parameters from a given data set. (For a review see, e.g., Ref.\citen{rf:TTH}.) 
The primary goal of the present paper is to determine the quality necessary for 
a data set from gravitational lensing systems in order to probe dark energy. 

A gravitational lensing system can be used to measure the ratio of 
(angular diameter) distances, while the supernovae can be used to determine 
the luminosity distance itself. One might ask whether the method of the gravitational 
lens provides new independent information. The second goal of the 
present paper is to answer this question. We show that 
the gravitational lens method can be a useful tool that complements 
the supernova method as a probe of dark energy.

This paper is organized as follows. In \S 2, we briefly review basic 
formulas. Then, we investigate the capability of the method employing 
strong gravitational lensing systems, focusing on errors in determining 
cosmological parameters, in \S 3. Uncertainties involved in modeling 
a lens galaxy are discussed in \S 4. In \S 5, we discuss 
the redshift distribution of a lensing system for the purpose of optimizing 
the method. Section 6 is devoted to a summary and conclusions. Throughout 
this paper we use units in which the velocity of light, $c$, equals $1$.

\section{Basic formulas}

In this section we give a brief review of basic formulas for 
the topics considered in the present paper. We restrict ourselves to a 
spatially flat FRW universe, in which case the Friedman equation is 
written $H(z)^2= H_0^2[\Omega_0 (1+z)^3+(1-\Omega_0)f(z)]$,
where $H_0=100h{\rm km/s/Mpc}$ is the Hubble constant, $\Omega_0$   
is the cosmic density parameter of the matter component, and 
$f(z)$ describes the redshift-evolution of the dark energy.
When the cosmic equation of state 
is parameterized as $w(z)=w_0(1+z)^\nu$, we have
\begin{equation}
  f(z)=(1+z)^3 \exp\biggl[3w_0\biggl({(1+z)^\nu-1\over \nu}\biggr)\biggr],
\end{equation}
which reduces to $f(z)=(1+z)^{3(1+w_0)}$ in the limit $\nu\rightarrow 0$. 
In a spatially flat universe, the angular diameter distance 
between $z_1$ and $z_2$ is 
\begin{eqnarray}
  D_A(z_1,z_2)={1\over H_0(1+z_2)}\int_{z_1}^{z_2}
  {dz'\over
  [\Omega_0(1+z')^3+(1-\Omega_0) f(z')]^{1/2}},
\label{DA}
\end{eqnarray}
where we have assumed $z_1<z_2$.

In order to estimate the accuracy to which we can constrain $\Omega_0$,
$w_0$, and $\nu$ with measurements using gravitational lensing 
systems, we employ the Fisher-matrix approach. 
With the Fisher-matrix analysis, one can estimate the best statistical 
errors on parameters from a given data set.
For this reason, this approach is widely used to estimate
how accurately cosmological parameters are determined from the large 
scale structure of galaxies or the cosmic microwave background 
anisotropies.\cite{rf:TTH,JKKSA,JKKSB} 
Tegmark et~al. applied the Fisher-matrix analysis to supernova data 
sets.\cite{rf:TEHK} Their analysis is useful for the present study.
In general, the Fisher-matrix is defined by
\begin{equation}
  \F_{ij}=\biggl<-{\partial^2 \ln L \over 
  \partial \theta_i\partial\theta_j}\biggr>,
\label{fisher}
\end{equation}
where $L$ is the probability distribution function of a data set, given 
model parameters $\theta_i$. 

\begin{figure}[b]
\begin{center}
    \leavevmode
    \epsfxsize=8cm
    \epsfbox[20 150 600 720]{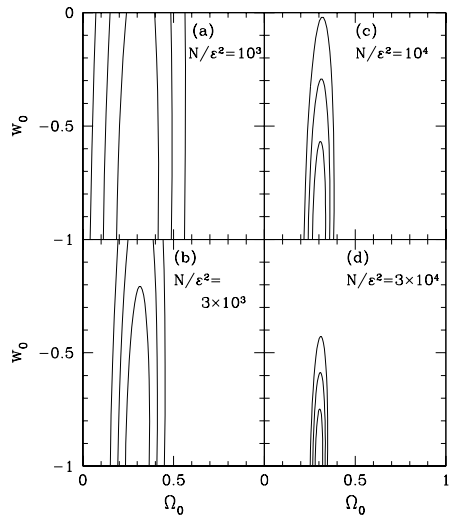}
\end{center}
\caption{Confidence regions from various data sets for
gravitational lensing systems. 
Here we have chosen the target parameters as $\Omega_0=0.3$ and $w_0=-1$, 
with the number and the individual uncertainties
as follows: (a) $N/\varepsilon^2=10^3$; (b) $N/\varepsilon^2=3\times 10^3$;
(c) $N/\varepsilon^2=10^4$; (d) $N/\varepsilon=3\times10^4$, where we have 
set $\varepsilon_n$ equal for all systems, $\varepsilon_n=\varepsilon$. 
The curves are the contours of the $68 \%$, $95 \%$, and $99.7 \%$ confidence 
regions. We chose the redshift distribution of lensing systems 
$z_{\it S}$ and $z_{\it L}$ through a homogeneous random process 
in the ranges $1\leq z_{\it S}\leq 3$ and $0.3\leq z_{\it L} 
\leq z_{\it S}-0.5$, respectively.}
\label{fa}
\end{figure}

To compute the Fisher matrix, we need to make an assumption 
concerning the probability function $L$, and more specifically
concerning the statistics of errors of measurements.
The method employing a strong gravitational lensing system 
relies on the assumption that the ratio of the angular 
diameter distances, $D_{\it LS}/D_{\it S}$, is precisely 
determined for each system, \cite{rf:FH,rf:FY} where 
$D_{\it LS}$ is the angular diameter distances between 
the lens and a source object, and $D_{\it S}$ is the 
distance between the source and the observer. 
A simple example in which $D_{\it LS}/D_{\it S}$ is determined
with a lensing system is considered in \S 4. 
Here, we assume that the lens model 
can be determined by measurements of images of a lensed 
object and the velocity dispersion of the lens galaxy, as well as 
the redshifts of the source and the lens objects. 
In a realistic situation, observational data is contaminated with 
errors in modeling the lensing potential and measurement 
of the velocity dispersion, as we discuss in \S 4. 
Hence, an observed value of $D_{\it LS}/D_{\it S}$ contains error. 
Throughout this paper, 
we also assume that this error can be described 
by a Gaussian random variable, for simplicity.
For definiteness we assume that a data set consists of $N$ gravitational 
lensing systems and that the observed ratio of the angular diameter 
distances of the $n$-th lens system can be expressed as
$R_n={D_{{\it LS}n}/ D_{{\it S}n}} + \Delta_n$,
where $D_{{\it LS}n}$ and $D_{{\it S}n}$ are the angular diameter distances 
$D_{\it LS}$ and  $D_{\it S}$ of the $n$-th lensing system, and 
the error $\Delta_n$ behaves as a Gaussian variable 
with zero mean, $\langle \Delta_n \rangle =0$. 
In this case the Fisher matrix (\ref{fisher}) can be written 
\begin{eqnarray}
  \F_{ij}&=&\sum_{n=1}^N 
        \biggl(2+{1\over \varepsilon_n^2}\biggr) 
                   W_{ni}W_{nj}
\nonumber
\\ 
        &\approx&  {1\over \varepsilon^2} \sum_{n=1}^N W_{ni}W_{nj},
\label{fish_form}
\end{eqnarray}
where 
$W_{ni}\equiv{\partial \ln \big(D_{{\it LS}n}/D_{{\it S}n}\big)/ \partial \theta_i}$,
$\varepsilon_n$ describes the variance of the Gaussian variable with
$\langle\Delta_n^2\rangle=(D_{{\it LS}n}/D_{{\it S}n})^2\varepsilon_n^2$,  and the
last approximate equality holds in the case that $\varepsilon_n(=\varepsilon)
\ll 1$ for all $n$. 
In contrast to the case of supernovae,\cite{rf:TEHK}
the factor 2 appears on the right-hand side of Eq.~(\ref{fish_form}). 
The corresponding term can be ignored in the case 
$\varepsilon\ll1$. This factor is traced back to the 
assumption that the error $\Delta_n$ is proportional to
 $D_{{\it LS}n}/D_{{\it S}n}$.
By using the Bayse theorem, the probability distribution in the 
parameter space can be written 
\begin{eqnarray}
 P(\theta_i)\propto \exp\biggl[-{1\over 2} \sum_{ij}(\theta_i-\theta_i^{tr})
  F_{ij}(\theta_j-\theta_j^{tr})\biggr],
\label{Prob}
\end{eqnarray}
where we have assumed that errors in the target model parameters 
$\theta_i^{tr}$ are small. Thus the Fisher matrix gives the 
uncertainties in the parameter spaces, which are described 
by a Gaussian distribution function around $\theta_i^{tr}$.

\section{Estimating errors in determining cosmological model parameters}

\begin{figure}[b]
\begin{center}
    \leavevmode
    \epsfxsize=6cm
    \epsfbox[20 150 600 720]{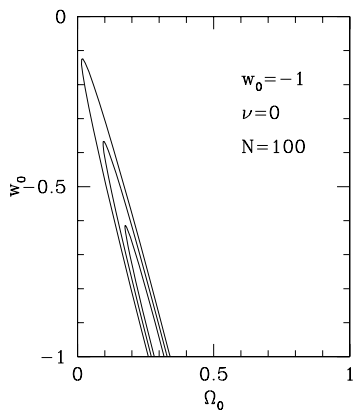}
\end{center}
\caption{Confidence regions from a data set from supernovae with 
individual uncertainties of $0.15$ magnitude. Here, $100$ 
supernovae randomly distributed in the range $0.5\leq z\leq 1.5$ 
are considered. The curves are the contours of $68 \%$, $95 \%$, 
and $99.7 \%$ confidence regions,
and the same target model as Fig.~1 is used.}
\label{fb}
\end{figure}

We now demonstrate how accurately cosmological model parameters 
can be determined by using the Fisher-matrix formalism. 
We first consider the model in which the cosmic equation of state 
is constant ($w(z)=w_0$), which is equivalent to the assumption 
$f(z)=(1+z)^{3(1+w_0)}$.
Considering the Fisher-matrix to be the $2\times2$ matrix corresponding to 
$\Omega_0$ and $w_0$, we show how accurately these two parameters can be 
determined.
Figure 1 displays the confidence regions in the $\Omega_0$-$w_0$ plane 
described by the probability function (\ref{Prob}),  where
we used the cosmological parameters $\Omega_0=0.3$ and $w_0=-1$ 
as the target model parameters. We considered various data 
sets for the number and the variance of the errors in
the gravitational lensing systems.  The quality of the data is 
characterized by $N/\varepsilon^2$, because the Fisher-matrix 
scales in proportion to $N/\varepsilon^2$, whose value is shown 
on each panel of the figure.
For the redshift distribution of the lensing systems, 
we chose $z_{\it S}$ and $z_{\it L}$ through a homogeneous random process in 
the ranges $1\leq z_{\it S}\leq 3$ and $0.3\leq z_{\it L} \leq z_{\it S}-0.5$.
The three curves in each panel indicate
the $68 \%$, $95 \%$, and $99.7 \%$ confidence regions. 
Figure 1 shows that the method of the gravitational lens
is sensitive to $\Omega_0$, but rather insensitive to $w_0$ 
in the case that $N/\varepsilon^2$ is small. It also shows
that a useful constraint on $\Omega_0$ and $w_0$ can be obtained 
if many lensing systems are measured precisely and $N/\varepsilon^2$ 
is sufficiently large.

\begin{figure}[t]
\begin{center}
    \leavevmode
    \epsfxsize=8cm
    \epsfbox[20 150 600 720]{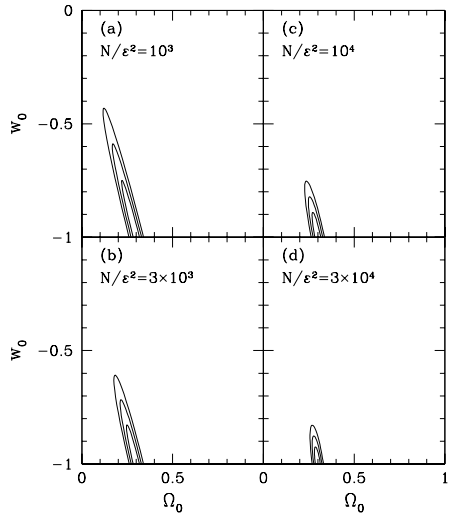}
\end{center}
\caption{Same as Fig.~1, but used combining the 
constraint from the data set of $100$ supernovae displayed in Fig.~2.}
\label{fc}
\end{figure}

One might think that the gravitational lens method 
can be no better than the supernova method  
because the latter supernova measures the luminosity distance itself, 
while the former measures only the ratio of the 
(angular diameter) distances.
However, in principle, these two methods give independent information 
and can be complementary.
Figure 2 demonstrates how accurately we can put a constraint on model 
parameters using a data set from $100$ supernovae randomly chosen in 
the range $0.5\leq z\leq 1.5$ with individual statistical uncertainties 
of $0.15$ mag.\cite{rf:TEHK} 
By comparing Figs.~1 and 2, it is clear that the two methods provide 
independent constraints. Thus, using both methods, we can eliminate
ambiguities in determining model parameters that arise when using 
only one. Figure 3 displays the confidence 
regions in the $\Omega_0$-$w_0$ plane, which is the same as  Figure 1, 
but with the addition of the constraints obtained from the data set of 
$100$ supernovae presented in Fig.~2. 
It is clear from Fig.~3 that the ability to constrain 
$\Omega_0$ and $w_0$ is significantly improved.

We next consider the case that the density parameter $\Omega_0$ 
is fixed from other experiments. We assume $\Omega_0=0.3$.
By considering the Fisher matrix to be the $2\times2$ matrix 
corresponding to the parameters $w_0$ and $\nu$, Fig.~4 
displays the confidence regions on the $w_0$-$\nu$ plane, where we 
used $w_0=-1$ and $\nu=0$ as the target model parameters. 
This figure displays the results using both the gravitational
lensing systems and the data set of $100$ supernovae, for which
we assumed the same range of distribution and uncertainty 
of sources as in Fig.~2. The constraint on $\nu$ might not 
be strict when $N/\varepsilon^2$ is of order $10^3$, but 
the useful constraint is obtained when $N/\varepsilon^2\sim O(10^4)$. 
Hence precise measurements of many gravitational lensing systems 
might provide a useful probe, even for the parameter $\nu$.

\begin{figure}[t]
\begin{center}
    \leavevmode%
    \epsfxsize=8cm 
    \epsfbox[20 150 600 720]{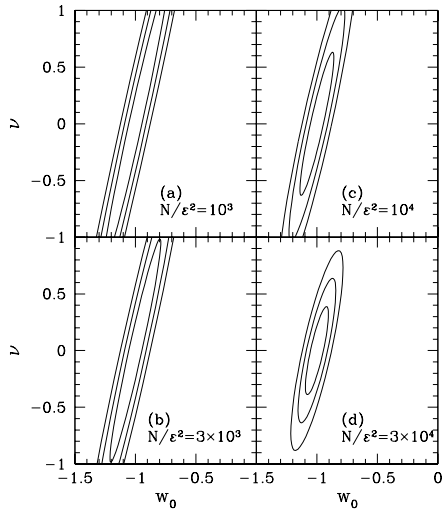}
\end{center}
\caption{Confidence regions in the $w_0$-$\nu$ plane, with
density parameter fixed as $\Omega_0=0.3$. 
The target model parameters
$w_0=-1$ and $\nu=0$ are used. The situation regarding 
the redshift distribution and $N/\varepsilon^2$ are the
same as in Fig.~1.}
\label{fd}
\end{figure}

\section{Errors in modeling lens galaxies}
In this section we discuss $\varepsilon$, which characterizes
systematic errors in the ratio $D_{\it LS}/D_{\it S}$. 
For this purpose, we start by demonstrating one of the  
simplest ways to determine the ratio, adopting a lens potential 
based on an isothermal ellipsoid model.\cite{rf:FY,KSB}
In this model, the lens equation gives an elliptical image 
of the Einstein ring with minor and major axes 
\begin{equation}
 \theta_{\pm} = \theta_E \sqrt{1 \pm \epsilon},
\label{DD}
\end{equation}
with
\begin{eqnarray}
  \theta_E={4\pi \sigma^2_v D_{\it LS}\over D_{\it S}},
\end{eqnarray}
where $\sigma_v$ is the one-dimensional velocity dispersion, and
the ratio $e$ of the minor axis to the major axis is related 
to the ellipticity $\epsilon$ by $e=\sqrt{(1+\epsilon)/(1-\epsilon)}$.
Thus the ratio $ D_{\it LS}/ D_{\it S}$ can be determined by 
measurements of $e$, $\theta_E$ and $\sigma_v$.

We note that several authors have recently pointed out 
possible errors involved in modeling the lensing 
potential.\cite{PBBZG,Golse,Kundson,CT} 
With regard to this matter, we first consider 
a possible density profile that differs from the 
singular isothermal profile, though we restrict
ourselves to a spherically symmetric density profile. 
We assume a generalized Navarro-Frenk-White\cite{NFWA,NFWB} 
or Zhao density profile,\cite{Zhao}
\begin{eqnarray}
  \rho(r)={\rho_s\over g(r/r_s)},
\end{eqnarray}
where we have defined 
\begin{eqnarray}
  g(x)=x^\alpha(1+x)^{\zeta-\alpha},
\end{eqnarray}
where $\alpha$ and $\zeta$ are parameters and $r_s$ is 
a characteristic length scale (cf. ref.~\citen{Faber}). 
In this case, the Einstein ring radius is determined by 
solving the lens equation
\begin{eqnarray}
  \theta_E-F(\theta_E,D_{\it L}/r_s,\alpha)4\pi\sigma_v^2
  {D_{\it LS}\over D_{\it S}}=0,
\label{lequation}
\end{eqnarray}
where we have defined 
\begin{eqnarray}
  F(\theta_E,D_{\it L}/r_s,\alpha)=
  {1\over \theta_E} \int_0^{\theta_E} d\theta' \theta'
  {1\over \pi}
  \int_{-\infty}^{\infty} dt 
  {D_{\it L}/r_s\over g\bigl(\sqrt{\theta'^2+t^2}D_{\it L}/r_s\bigr)}.
\end{eqnarray}
Here we have considered the case $\zeta=2$, so that the lens model 
possesses the same velocity dispersion as the singular 
isothermal sphere at large distances. In the limit 
$\alpha\rightarrow2$,
we have $F(\theta_E,D_{\it L}/r_s,\alpha)=1$, and the model
reduces to the singular isothermal model.
This model indicates how the ratio $D_{\it LS}/D_{\it S}$,
which is determined by solving the lens equation (\ref{lequation}),
depends on the density profile parameterized 
by $r_s$ and $\alpha$ with the same observable quantities 
$\theta_E$ and $\sigma_v^2$.
Figure 5 plots the normalized difference between $D_{\it LS}/D_{\it S}$ 
and the corresponding value in the singular isothermal model.
Explicitly, this figure displays contours of the quantity
\begin{eqnarray}
  {D_{\it LS}/D_{\it S} -D_{\it LS}/D_{\it S}\vert_{SIS}
  \over D_{\it LS}/D_{\it S}\vert_{SIS}}
  ={1\over F(\theta_E,D_{\it L}/r_s,\alpha)}-1,
\end{eqnarray}
in the $r_s$-$\alpha$ plane, 
where $D_{\it LS}/D_{\it S}\vert_{SIS}=\theta_E/(4\pi\sigma_v^2)$ 
is the value of this ratio in the singular isothermal model. 
In this figure we fixed $\theta_E=1''$ and used the cosmological 
model with density parameters $\Omega_0=0.3$ and $w(z)=-1$. 
The four panels correspond 
to the cases in which the redshift of the lens is $z_{\it L}=0.5,~1.0,~1.5$
and $2.0$. For example, the normalized difference
is of order $5\%$ and $10\%$, respectively, for  $\alpha=1$ 
and $\alpha=0$ at $r_s=0.1~h^{-1}{\rm kpc}$,
which depends weakly on the redshift of the lens objects.
The difference between the ratios becomes larger as $r_s$ increases
or $\alpha$ decreases. We thus find that the ambiguity in the density profile 
can be an important factor in modeling the lensing potential.

\begin{figure}
\begin{center}
    \leavevmode
    \epsfxsize=8cm
    \epsfbox[20 50 650 810]{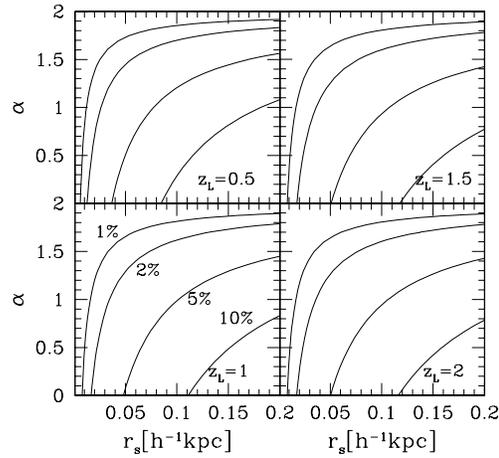}
\end{center}
\caption{Contours of the normalized difference defined by Eq.~(4.7) in 
the $r_s$-$\alpha$ plane. This figure displays the normalized difference
between the ratio $D_{\it LS}/D_{\it S}$ and the corresponding value for
the singular isothermal model. Note that $\alpha=2$ represents the case of 
the singular isothermal model. We fixed $\theta_E=1''$ and adopted 
the cosmological model with $\Omega_0=0.3$ and $w(z)=-1$. In each panel, 
the redshift of the lens object is adopted as $z_{\it L}=0.5,~1.0,~1.5$
and $2.0$, respectively. The curves are the contours of the 
$1\%$,~$2\%$,~$5\%$ 
and $10\%$ levels from left to right for all panels.}
\label{fig_erro}
\end{figure}

We next consider the errors involved in the observation of the 
velocity dispersion, which we have not yet discussed.
Mortlock and Webster systematically investigated the 
errors involved in gravitational lensing statistics and pointed 
out that the greatest uncertainty comes from the dynamical 
normalization of lens galaxies.\cite{Mortlock}
According to their results, the velocity dispersion 
$\sigma_v$ in the isothermal sphere model (the line-of-sight 
velocity dispersion away from core radius) can be related 
with the observed line-of-sight velocity dispersion of 
stars $\sigma_{||}(R_f)$, assuming a circular aperture of 
radius $R_f$ as
\begin{equation}
  \sigma_v=\sigma_{||}(R_f)\biggl(A+B{r_c\over R_g}\biggr),
\end{equation}
where $A$ and $B$ are constants independent of the core radius $r_c$,
and $R_g$ is the effective radius of the lens galaxy. Kochanek
used $A=1$ and $B=2$.\cite{Kochanek} Then, the non-dimensional 
uncertainty of $\sigma_v$ may be expressed as
\begin{equation}
  {\delta\sigma_v\over\sigma_v}
  \simeq
  {\delta\sigma_{||}(R_f)\over\sigma_{||}(R_f)}
  +{B \delta r_c /R_g \over A+Br_c/R_g},
\label{deltasigma}
\end{equation}
where $\delta\sigma_{||}(R_f)$ and $\delta r_c$ represent
uncertainties on $\sigma_{||}(R_f)$ and $r_c$. Because the 
ratio of the distance $D_{\it LS}/D_{\it S}$ is proportional
to $\sigma_v^2$, the error that contributes to $\varepsilon$ is 
estimated as ${\varepsilon} \simeq 2\delta\sigma_v/\sigma_v$,
where we have assumed $\delta\sigma_v/\sigma_v\ll1$.
Concerning the uncertainty of the steller velocity dispersion, 
measurements of the velocity dispersion of lensing galaxies 
with accuracy to within approximately $3$\% have been made 
with the Keck-II $10$m telescope,\cite{TonryII,Tonry}
while the uncertainty on the core radius $r_c$ might be 
very important. For example, the second term on
the right-hand side of (\ref{deltasigma}) is of order
$0.1$ for $r_c=0.1$~kpc and $R_g=2$~kpc. Thus the investigation
of this section demonstrates the importance of precisely modeling the 
lensing galaxy, taking the possible finite core radius and
the density profile into account in a realistic analysis. 

\section{Optimizing constraint on model parameters}
In this section we address the problem of determining the optimal 
distribution in redshift of gravitational lensing systems 
in order to best constrain the cosmological parameters.
This problem might be only of theoretical interest,
but having such information could be useful in planning observation 
of gravitational lensing systems. 
The same problem has been considered by Huterer and 
Turner with regard to supernova data.\cite{HTb} Our discussion here
is based on the following observation: By a theorem concerning 
the Fisher matrix, it can be shown that the minimum error 
attainable on $\theta_i$ is expressed by the diagonal part of 
the inverse Fisher-matrix if the other parameters are known.\cite{rf:TTH}
To address the above stated problem, we 
investigate what the pair of redshifts of the source and the 
lens objects maximize the diagonal part 
\begin{equation}
 \varepsilon^2 F_{ii} \simeq \frac1{D_{\it S}^2 D_{\it LS}^2} 
  \left( D_{\it S} \frac{\partial D_{\it LS}}{\partial \theta_i}
       - D_{\it LS} \frac{\partial D_{\it S}}{\partial \theta_i} \right)^2,
\label{l-9}
\end{equation} 
where the case of one lensing system is assumed for simplicity.
It is useful to note how the normalized variation of the ratio
$D_{\it LS}/D_{\it S}$
is related with the variances of the cosmological parameters and
the Fisher-matrix:
\begin{eqnarray}
  {\Delta(D_{\it LS}/D_{\it S})\over D_{\it LS}/D_{\it S}}
 =\sum_{i}W_i\Delta \theta_i
 =\varepsilon F_{\Omega_0\Omega_0}^{1/2}\Delta\Omega_0
 +\varepsilon F_{w_0w_0}^{1/2}\Delta w_0
 +\varepsilon F_{\nu\nu}^{1/2}\Delta \nu.
\end{eqnarray}

\begin{figure}[t]
\begin{center}
    \leavevmode
    \epsfxsize=12cm
    \epsfbox[20 50 650 810]{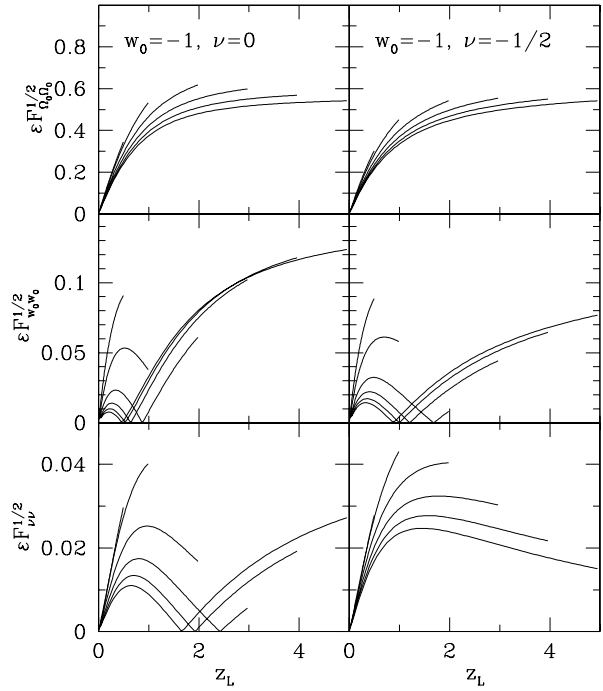}
\end{center}
\caption{Amplitude of $\varepsilon F_{ii}^{1/2}$ as a function of 
the redshift of the lens object $z_{\it L}$ with various redshifts of
the source object $z_{\it S}$. The top, middle and bottom panels, 
respectively, display the components $\Omega_0$, $w_0$ and $\nu$.
The left and right panels correspond to the case of the target
model with $\Omega_0=0.3$, $w_0=-1$ and $\nu=0$ and  with 
$\Omega_0=0.3$, $w_0=-1$ and $\nu=-0.5$, respectively. 
In each panel, the curves correspond to $z_{\it S}=0.5,~1,~2,~3,~4$ and 
$5$, from top to bottom near $z_{\it L}=0$.}
\label{fig_omega}
\end{figure}

Figure 6 displays the amplitude of the diagonal part 
given in $\varepsilon F_{ii}^{1/2}$ 
as a function of $z_{\it L}$ for 
$z_{\it S}=5,~4,~3,~2,~1$ and $0.5$. The left and right 
panels depict the cases of the target model with 
$\Omega_0=0.3$, $w_0=-1$ and $\nu=0$ and with 
$\Omega_0=0.3$, $w_0=-1$ and $\nu=-1/2$, respectively.
The top, middle, and bottom panels, respectively, display the components
$\theta_i=\Omega_0$, $w_0$, and $\nu$. The amplitude shown in each 
panel confirms that $\Omega_0$ can be most easily determined, while $w_0$ 
and $\nu$ are relatively difficult to determine than $\Omega_0$.
From the top panels, it is clear that $\Omega_0$ is more precisely 
determined as $z_{\it L}$ becomes larger and that the precision with
which $\Omega_0$ can be determined is almost the same for all 
$z_{\it L}$ satisfying $z_{\it S}> z_{\it L} \simgt 1$. 
Contrastingly, the situation for $w_0$ and $\nu$ is complicated. 
We see that as the redshifts of the source and the lens objects become 
larger 
(e.g., for $z_{\it S}> z_{\it L} \simgt 2$), the precision with which
$w_0$ and $\nu$ can be determined becomes better.
However, this is not always the case, because $F_{ii}$ becomes zero 
for certain combinations of $z_{\it L}$ and $z_{\it S}$, which depend 
on the target model parameters. From the middle and lower panels, we can 
find another peak for $z_{\it S}\simlt 1$ at which $w_0$ and $\nu$ can be
determined precisely. It is thus seen that low redshift lensing 
systems are also useful to determine the redshift evolution 
of dark energy. This is consistent with the result of
Ref.~\citen{rf:YF}, and it is also consistent with the result obtained 
from supernova analysis.\cite{HTb} This feature can be understood from 
the fact that the dark energy component increases relative to the 
matter component as the redshift decreases.

\section{Summary and conclusions}
In summary, we have assessed the feasibility of probing dark energy
using gravitational lensing systems. We have demonstrated how
precisely the model parameters can be constrained using the 
Fisher-matrix formalism. Our results show that the method 
might place useful constraints on the density parameters and 
the cosmic equation of state if many lensing 
systems are measured precisely. Our results also demonstrate the 
toughness of the method. It was found that $N/\varepsilon^2 \simgt 10^3$ 
is required to probe the cosmic equation of state $w_0$ 
even if combined with the supernova method employing data from 100 
supernovae, and $N/\varepsilon^2\simgt 10^4$ is required to determine 
$w_0$ using only gravitational lensing systems.

In order to realize the condition $N/\varepsilon^2=10^3$ -- $10^4$ 
we need $\varepsilon \simlt 0.1$ -- $0.3$ for $N=100$.
This required accuracy of measurement and modeling the lensing 
system might be difficult to realize, due to theoretical 
and observational ambiguities.\cite{PG}
As discussed in \S 4, uncertainties involving the density 
profile and the velocity dispersion are problematic for 
precise modeling of a lensing system.
Nevertheless, the possibility of precisely modeling lens potentials 
has been studied. For example, the ability of observed 
Einstein rings to precisely model lens potentials has been
discussed by several authors.\cite{KKM,CKMK}
The advantage of Einstein rings is the richness of the
information they provide. The authors of those works have claimed 
that the distribution
of the magnifications of Einstein rings makes it possible
to eliminate lens model ambiguities and to model the mass 
distribution of a lens precisely.\cite{KKM}
Furthermore, they have pointed out that Einstein rings 
must be found frequently in multiply imaged 
quasars with sufficiently long observations, because it is 
likely that all quasars and AGN have host galaxies.
Thus we might expect that studies of modeling lensing 
galaxies will provide more accurate results in the near future, 
employing many sample galaxies in SDSS, and that 
future progress might allow us to solve the problems involved in
modeling the lensing potential.\cite{Rix} 

In \S 5, we considered optimizing the analysis. Though 
our consideration was restricted to the case of one gravitational 
lensing system, it is worth comment.
We found that the gravitational lensing system method is most 
sensitive to $\Omega_0$, and less sensitive to $w_0$ 
and $\nu$. To determine $\Omega_0$, it is most advantageous 
to have larger redshifts of the source and the lens objects,
while this is not always true for $w_0$ and $\nu$. This feature 
is due to the fact that the dark energy component decreases 
relative to the matter component 
as the redshift increases. Lensing systems with smaller 
redshifts can be useful to probe dark energy.

\section*{\bf Acknowledgements}
We are grateful 
to Y.~Kojima and S.~Yoshida for useful conversations and comments. 
We also thank T.~Hamana for his important comments and discussions.
Parts of \S 4 are based on the valuable 
communications with him. We also thank R.~R.~Caldwell for useful conversations
and important comments on a preliminary manuscript, which helped improve
the paper. We are also grateful to referees for helpful comments.
One of the authors (K.~Y.) thanks Professor S.~D.~M.~White and the 
people at the Max-Planck Institute for Astrophysics (MPA) 
for their hospitality. Some 
parts of this work were done there. He acknowledges financial support
from the Deutscher Akademischer Austauschdienst (DAAD) 
and the Japan Society for the Promotion Science (JSPS) for 
a visiting program to MPA. 
This research was also supported by the Inamori Foundation.

\end{document}